\documentclass[aps,prx,twocolumn,floats,showpacs,superscriptaddress,nofootinbib]{revtex4-1}
\usepackage{graphicx,epsfig}
\usepackage{times,bbm}
\usepackage{graphics,dcolumn,bm,float}
\usepackage{amssymb,amsmath,rotate,color,amsfonts}
\usepackage[title,titletoc,toc]{appendix}
\usepackage{mathtools}
\usepackage{booktabs}
\usepackage{tcolorbox}
\usepackage[pagebackref=false,colorlinks,linkcolor=magenta,citecolor=blue,urlcolor=magenta]{hyperref}

\usepackage{wrapfig}
\usepackage{cancel}
\usepackage{lipsum}
\usepackage{mwe}
\usepackage[mathlines]{lineno}
\usepackage[mathscr]{euscript}
\usepackage{hyperref}
\usepackage{breqn}
\usepackage{bbold}
\usepackage{pgfplots}
\usepackage{float}
\usepackage{tkz-euclide}
\usepackage{braket}
\usepackage{physics}
\usepackage[caption=false]{subfig}
\usepackage[export]{adjustbox}
\usepackage{tikz}
\usepackage{slashed}
\usepackage{bm}
\usetikzlibrary{through,calc}
\usetikzlibrary{positioning}

\newcommand{\be}{\begin{equation}}
\newcommand{\ee}{\end{equation}}

\newcommand{\bea}{\begin{eqnarray}}
\newcommand{\eea}{\end{eqnarray}}

\newcommand{\bmt}{\left[\begin{matrix}}
\newcommand{\emt}{\end{matrix}\right]}

\begin{document}
\preprint{}
\title{Material and size dependent corrections to conductance quantization in anomalous Hall effect from anomaly inflow}
\author{Armin Ghazi}
\affiliation{Department of Physics$,$ Sharif University of  Technology$,$ Tehran 11155-9161$,$ Iran}
\author{S. A. Jafari}
\email{Current address: 2nd Institute of Physics C, RWTH Aachen University, Aachen  52074, Germany. Email: akbar.jafari@rwth-aachen.de}
\affiliation{Department of Physics$,$ Sharif University of  Technology$,$ Tehran 11155-9161$,$ Iran}
\date{\today}

\begin{abstract}
In quantum anomalous Hall (QAH) systems, the Hall conductance is quantized and the corresponding effective topological theory of the system is the Chern-Simons theory. The conductance quantum is given by the universal constant  $e^2/h$ -- the inverse von Klitzing constant -- that is independent of the bulk gap, as well as the size of the system.  This picture relies on the assumption that the edge modes are sharply localized at the edge, i.e. they have zero width. We show that considering the physical case where the edge modes have finite localization length $b$, the effective action would not be topological in bulk direction anymore. Due to non-zero $b$ the conductance quantum will be corrected as $(1-\varepsilon)e^2/h$ where $\varepsilon$ encompasses the non-universal (i.e. material/sample dependent) part that is determined by the dimensionless ratios $\frac{gb}{\hbar v_F}$ and $\frac{b}{L}$ where $g,v_F,L$ are the bulk gap, Fermi velocity and sample length. To compute the non-universal correction $\varepsilon$ we use anomaly inflow framework according to which the bulk action produces the correct amount of anomaly inflow that would cancel the anomaly of the chiral edge modes. These corrections place limits on the precision of measurable quantization in units of the inverse von Klitzing constant for QAH systems with smaller sizes and/or smaller bulk gaps. Our result suggests that the failure of precision measurements to reproduce the exact conductance quantum $e^2/h$ is not an annoying sample quality issue, but it contains the quantitative physics of anomaly inflow that can be inferred by the systematic study of such corrections.  
\end{abstract}



\maketitle
\narrowtext

\section{Introduction}

The hallmark of symmetry protected topological (SPT) phases is their gapless zero modes on their boundary \cite{PhysRevB.87.155114}. in the fermionic SPTs,  When the gapped fermions are integrated out, effective electromagnetic theory in SPT phases realizes a topological term that represents the topological nature of the material. Such topological terms are universal in the sense that they are independent of the bulk gap and the system size and have an intimate relation with the edge modes via anomaly inflow \cite{callan_anomalies_1985}. This intimate relation would force the bulk action to correctly manifest the features of edge modes. This relation between the topology of the bulk and the existence of zero modes on the boundary is known as bulk-edge correspondence. Is there a more quantitative expression of this correspondence to allow for a measurement that would shed more light on this correspondence?

The effective action derived by integrating out the bulk gapped modes assumes that the edge modes live on an exact one lower dimensional space, i.e. they are sharply localized in the boundary without any spatial profile along the bulk direction. Such a Dirac-delta localization of zero modes at the boundary is a convenient assumption to formalize the correspondence between the bulk and the edge. However, in actual topological materials the localization length can not be less than the atomic scales. Assuming that the edge modes are localized within the range $b$ of the boundary, introduces a length scale to the bulk-edge correspondence. It would be natural to expect that the spatial profile of zero modes contribute corrections to the effective action of the gapped bulk. One naively anticipates that the localization length $b$ of the zero modes will be combined with the length $L$ representing the system size to form a dimensionless ratio $b/L$ that will ultimately control the possible corrections to the universal $b=0$ limit. We find surprisingly that as soon as $b$ is non-zero, it triggers dependence on both the sample size $L$ and the bulk gap $g$ via the dimensionless parameter $\frac{gb}{\hbar v_F}$. In this way, the effective action is not strictly topological in the sense that it will acquire non-universal corrections that are given by sample/material specific parameters. The bright side of such dependence is that its systematic study can be an experimental framework for the quantitative study of the anomaly inflow on the tabletop. 

In the $b=0$ universal limit, all the information that is matched between bulk and boundary via anomaly is captured by anomaly descent equations that relate three parties: anomaly in $d$ dimensions, Chern-Simons theory in d+1 dimensions, and topological characteristic classes in $d+2$ dimensions \cite{ALVAREZGAUME1985423}. But we show that \emph{in addition to this universal topological information, there is also non-universal correspondence between bulk and edge} which is captured by a deformation of the original descent equation containing the embedding information of $d$ dimensional anomalous theory inside the $d+1$ dimensional manifold. Rephrasing the above point in the context of QAH system, when the bulk theory has a non-trivial topology, the edge is characterized by exact $e^2/h$ conductance. Deviations of the bulk theory from topological theory (in bulk direction) will be reflected as corresponding deviations from the universal conductance $e^2/h$. Anomaly inflow encompasses both situations and therefore provides a framework to compute such corrections.

To demonstrate this phenomenon within the framework of anomaly inflow, we choose the QAH Effect (QAHE)~\cite{Liu2016}. The first step in promoting the classical version of the anomalous Hall effect~\cite{Nagaosa2006,Nagaosa2010} was a theoretical effort by Haldane proposing a model based on the honeycomb lattice that would give a non-zero Hall conductance $\sigma_{H}$ without requiring a magnetic field~\cite{Haldane1988}. ‌But the actual realization of QAHE is based on the surface Dirac cones of topological insulators. These Dirac cones can be gaped out by doping magnetic atoms offering a 2D massive Dirac cone~\cite{Yu2010} required for the realization of QAHE. These massive Dirac fermions will play the role of massive bulk Dirac fermions for the QAH transport at the edges as realized in recent experiments~\cite{Chang2013,Feng2015}. 

One important advantage of this experimental platform is that the bulk gap is caused by the external magnetic influence of the doped magnetic atoms. Therefore using external magnetic fields to force them away from the perpendicular direction~\cite{Kandala2015} can be thought of as a knob to tune the gap of the massive Dirac fermions. In this way, we propose the QAH transport in magnetically doped topological insulators as a suitable platform to study the anomaly inflow physics systematically by studying the size $L$ and gap $g$ dependence of the anomalous Hall conductance.  We model the QAHE with massive Dirac fermions in two space dimension coupled with a scalar field $\phi$ featuring a kink to which edge modes responsible for the QAHE are bound. The question we study in this work is the impact of the width of the kink that also represents the localization length of the edge modes, on the Hall conductance $\sigma_{\rm H}$ of the QAHE. 

The organization of the paper is as follows: In section~\ref{anomaly.sec} we describe how the anomaly inflow can be used to incorporate the localization size of the edge modes as corrections to universal conductance. In section~\ref{twoplus1d.sec} we implement the anomaly inflow computation for the 2+1 dimensional massive Dirac fermions of the QAHE. In section~\ref{corrections.sec} we compute the non-universal corrections arising from the finite localization size of the edge modes for a Sigmoid shaped mass profile the numerical results of which are presented in section~\ref{results.sec}. We end the paper with summary and discussions in section~\ref{discussions.sec}. 

\section{Anomaly inflow framework}
\label{anomaly.sec}
The non-trivial topology in topological materials manifests itself when boundaries or defects are introduced into such gapped systems. Then the topology will be represented by gapless excitations living on the interface which are protected by the quantum anomaly associated with these boundary zero modes. The phenomenon of canceling the gauge non-invariance of zero modes living on the boundary by the topological field theory of the bulk is known as anomaly inflow and was first introduced by Callan and Harvey~\cite{callan_anomalies_1985} following which there has been a plethora of examples of such bulk edge correspondence~\cite{Wen_2013,PhysRevB.43.11025,PhysRevB.87.155114,PhysRevB.85.045104,PhysRevB.87.134519}.

Considering a $d$-dimensional theory with global symmetry described by the group G. The symmetry would transform the degrees of freedom of system $\phi$ to $\phi_g$. 
If we promote such global symmetry to a local gauge symmetry and couple the system with background gauge field $A_\mu$ associated with this gauge symmetry, 
one can write the partition function of such theory in the background of this gauge field as
\begin{equation}
Z^d[A]=\int \mathcal{D}\phi e^{i S[\phi,A]},
\end{equation}
where $S[\phi, A]$ is the action of theory describing the dynamics of system in the presence of background gauge field A. Anomaly appears when such partition function does not remain invariant under gauge transformation of the gauge field, transforming $A$ to $A^g$, where g denotes parameter of local gauge transformation:
\begin{equation}
Z^d[A^g] \neq Z^d[A].
\end{equation}
The amount of non-invariance would depend generally on both A and g and can be classified by a phase factor $Z^d_{\rm anomaly}[A,g]$ defined as
\begin{equation}
Z^d[A^g] = Z^d[A].Z^d_{\rm anomaly}[A,g].
\label{eq anomaly def}
\end{equation}

The anomaly in $d$-dimension is intimately tied to the SPT phases in d+1 dimensions. 
There would always be an SPT phase living on a d+1 dimension manifold with boundary that produces the same amount of non-gauge invariance as $d$-dimensional anomalous system would produce. Therefore they both have the same classification~\cite{kapustin_fermionic_2015,Yonekura_2019}.

On the quantum materials side, one of the conceptually simple, but experimentally challenging phenomena was the observation of Quantum Anomalous Hall Effect (QAHE)~\cite{Chang2013} which relies on magnetic topological insulators~\cite{Tokura2019}. In this paper, following Harvey~\cite{Harvey_2001}, for the QAHE problem, we investigate how the spatial profile of the gapless boundary excitations in the bulk direction influences the effective action of the topological theory of the bulk. We obtain such dependence using the anomaly inflow for an anomalous Hall system. We discuss implications for the recent experiments on the QAHE.

\section{2+1 dimensional anomalous Hall system}
\label{twoplus1d.sec}
\begin{figure}[b] 
	\centering
	\includegraphics[width=80mm]{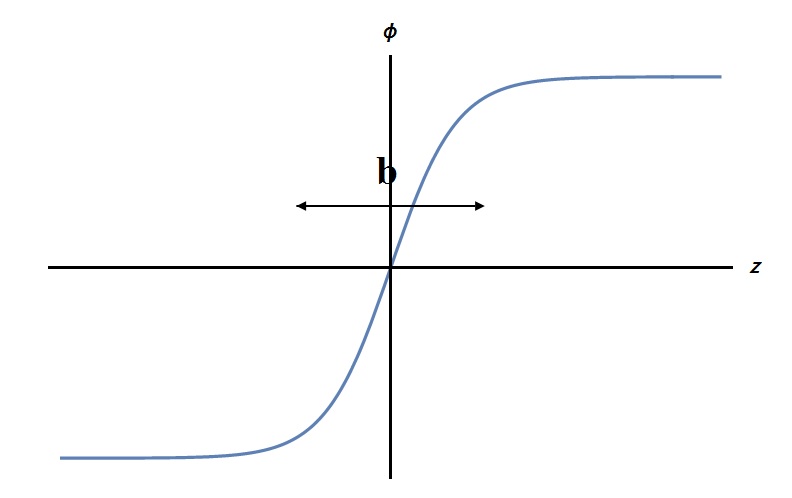}
	\caption{The domain wall profile}
	\label{wall1}
\end{figure}

Let us begin by considering a quantum anomalous Hall system in 2+1 dimension. Such a system can be described in low energy by a pair of Gapped Dirac theories \cite{Haldane1988} which due to the presence of mass term would break time reversal invariance. Each Dirac theory would contribute half quantized value $\frac{e^2}{2h}$ to integer quantized Hall conductance. Our starting point would be to consider one of these 2+1 dimensional Dirac equations in the background of a 1+1-dimensional domain wall. 
The domain walls can be produced in the process of symmetry breaking of a $Z_2$ symmetry or can be considered as an existing interface between two regions with different masses.
In the symmetry breaking case one can consider the Dirac equation interacting with a scalar $\phi(x)$ whose  Lagrangian is
\begin{equation}\label{lagrangian}
L= \bar{\psi}(x) i \gamma^\mu (D_\mu) \psi(x)+\frac{1}{2}\partial_\mu \phi \partial^\mu \phi-g \bar{\psi} \phi \psi(x) -V(\phi),
\end{equation}
where $g$ is the coupling between fermions and the background scalar field $\phi$ and has the dimension of mass (energy). The potential $V(\phi)$ induces $Z_2$ symmetry breaking and will produce a vacuum with either of the expectation values  $\langle  \phi \rangle=v$ or $\langle  \phi \rangle=-v$.  But there can also be a domain-wall solution such that $\phi$ interpolates between such two vacua with an asymptotic behavior specified by $\phi(-\infty)=-v$ and $\phi(+\infty)=v$. An example of such solutions is portrayed in Fig.~\ref{wall1}.

Such codimension q defects are known to exist when the broken symmetry G has non-trivial $q-1 $ 
homotopy groups $\pi_{q-1}(G)$ \cite{weinberg_2001}. Here we have codimension one defect or domain wall because $\pi_0(Z_2)=Z_2$.
In the presence of a domain wall, the Dirac fermions obey the equation of motion:
\begin{equation}
i \gamma^\mu \partial_\mu \psi(x)-g \phi(x) \psi(x)=0,
\end{equation}
and separating directions along the wall $a=0,1$ from  perpendicular spatial direction  $\mu=2$ or $z$, we will have 
\begin{equation}
i \gamma^a \partial_a \psi(x)+i\gamma^2\partial_2\psi-g \phi \psi=0,
\end{equation}
where $\gamma^\mu$ are gamma matrices obeying the Clifford algebra $\{\gamma^\mu,\gamma^\nu\}=2\eta^{\mu\nu}$ and $\eta^{\mu\nu}$ is the Minkowski metric of the space-time.

Considering the  profile of domain wall in the perpendicular direction to the wall, one can separate the Dirac equation into two parts using the ansatz,
\begin{equation}
\psi=\eta(x^a) \mathcal{F}(x_{\bot})=\mathcal{A}  \eta(x^a) e^{g\alpha(z)},
\end{equation}
where $a=0,1$ are the directions along the wall, $\eta(x^a)$ is the 1+1 dimensional spinor living on the wall and $\mathcal{F}(x_{\bot})= \mathcal{A} e^{g \alpha(z)}$ is the fermion profile in the direction normal to the wall with a normalization constant $\mathcal{A}$. Using this ansatz, we obtain an effective 1+1 D Dirac equation plus a constraint:
\begin{eqnarray}\label{dirac2d}
& i \gamma^a \partial_a \eta(x^a)=0,\\
& i(\gamma^z\eta ) \partial_z \alpha= \phi\ \eta.
\end{eqnarray}
The constraint can be solved if we have
\begin{equation}
i\gamma^z\eta = \pm \eta, \qquad \partial_z \alpha=\pm\phi.
\end{equation}
The second equation gives
\begin{equation}
\alpha(z)= \pm \int_0^z \phi(z^\prime) dz^\prime,
\end{equation}
from which the solution bounded to the wall will be
\begin{equation}
\psi(x)=\mathcal{A} \eta(x^b) e^{\pm g \alpha(z)}=\mathcal{A} \eta(x^b) e^{\pm  \int_0^z g \phi(z^\prime) dz^\prime}.
\end{equation}
Note that $i \gamma^z$ is the chirality operator in 1+1 dimension. So one implication of the constraint equation is that the 2D spinor has a definite chirality.  The other implication is that the positive chirality solution is exponentially growing in the direction normal to the wall and the negative chirality one is exponentially decaying in the same direction or in another word the negative chirality solution is localized on the domain wall.

Solving the reduced Dirac equation along the wall, Eq~\eqref{dirac2d}, one obtains the free propagating solution with momentum $p$,
\begin{equation}
\eta_p^\pm(t,x) = u_{\pm}  e^{-ip(t\pm x)},
\end{equation}
where $u_{\pm}$ is 2D spinor with definite chirality. Such  2D spinor is written as a subspace of a 3D spinor  

\begin{equation}
u_{\pm} = \frac{1}{\sqrt{2E}} \begin{pmatrix}
\sigma^\mu p_\mu \,\,\xi_\pm \\\
\bar{\sigma}^\mu p_\mu \,\xi_\pm
\end{pmatrix} ,
\end{equation}
where $\sigma^\mu=(1,\sigma^i)$, $\bar{\sigma}^\mu=(1,-\sigma^i)$ in which $\sigma^i$ are Pauli matrices and $\xi_\pm$ are chosen in a basis where $\xi_\pm=
\begin{pmatrix}
1\\\mp i \end{pmatrix}$. The full solution would be
\begin{equation}
\psi_p(x)=\mathcal{F}(x_{\bot}) u_{\pm}  e^{-ip(t\pm x)}=\mathcal{A} u_{\pm}  e^{-ip(t\pm x)}e^{\pm  \int_0^z g \phi(z^\prime) dz^\prime}.
\end{equation}
To be normalizable it must satisfy the normalization condition
\begin{equation}\label{norm}
\int d^2x \, \psi_p(x)^\dagger \psi_{p^\prime}(x)=2\pi \delta(p-p^\prime).
\end{equation}
This condition excludes the exponential growing solution from the list of viable solutions and also fixes the normalization constant $\mathcal{A}$ for the localized solution. Simplifying the expression by using the plane-wave solution along the domain wall we obtain the condition only for the transverse directions
\begin{equation}\label{normalization1}
\int dx_{\perp} \,\mathcal{F}^2(x_\bot)=1.
\end{equation}

\subsection{Anomaly on 1+1D domain wall}
Our goal here is to find the effective action of the gapped modes of the bulk in terms of the background $U(1)$ fields such that correctly portrays the anomaly inflow picture in the presence of edge modes with finite size localizations. To do so, we need to find the anomaly or non-invariance of effective action of zero mode fermions. Once we have their contribution to the anomaly, we will construct an effective action of the bulk such that the bulk effective action cancels the anomaly of localized zero modes.
If we were to consider the domain wall as an exact two-dimensional surface (not a 2D manifold embedded in 3D bulk), then one could compute the anomaly of 2D chiral fermions on it, which would produce the well known result
\begin{equation}
\partial_a j^a=\frac{-e}{4\pi} F= \frac{-e}{8\pi} \epsilon^{a b} F_{a b},
\end{equation}
which is the consistent $U(1)$ anomaly of gapless chiral fermions~\cite{Naculich1988}. There would be another concept known as Covariant anomaly, related to consistent anomaly but differs from it by a factor of two \cite{BARDEEN1984421}. Such non-conservation of $U(1)$ current $j^a$ is an immediate consequence of 
equation (\ref{eq anomaly def}) for the case of $U(1)$ background gauge fields ($A_\mu$). In this case we will have
\begin{equation}\label{eq anomaly U(1)}
Z[A+\frac{1}{e}d\lambda] = e^{i\int \frac{\lambda F}{4\pi} }Z[A].
\end{equation}
so here $Z_{\rm anomaly}[A,g]=e^{i\int \frac{\lambda F}{4\pi}}$, which is associated with the 3D Chern-Simons theory as its corresponding bulk theory. Now for the quasi-two dimensional anomaly produced along the domain wall, one should compute the contribution of zero modes which are localized along the domain wall in the original 3D theory.

\begin{figure}[t] 
	\centering
	\includegraphics[width=50mm]{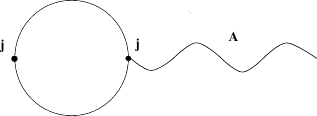}
	\caption{Zero mode Feynman diagram.}
	\label{screening}
\end{figure}

We can decompose the fermion field into two parts, the zero mode and non-zero modes.
\begin{equation}\label{decomp1}
\psi= \psi_{\rm zm}+ \psi_{\rm nzm},
\end{equation}
One can also assume the same splitting for Green's function
\begin{equation}\label{decomp2}
G(x_2,x_1)=\langle \psi(x_2)\psi(x_1)^\dagger \rangle= G(x_2,x_1)_{\rm zm}+G(x_2,x_1)_{\rm nzm}.
\end{equation}
To see the inflow picture, one should compute the U(1) current in this domain wall background. The total current should be conserved as the consequence of non-anomalous U(1) symmetry in the system. But as we know the current along the domain wall is non-conserved due to the anomaly, so one should see this non-invariance by computation of zero mode current associated with the Feynman diagram of Fig.~\ref{screening}.

Considering the partition function of the full fermionic theory in the background of U(1) gauge field $A_\mu$,
\begin{equation}
Z_T[A]=\int \mathcal{D}\bar{\psi}\mathcal{D}\psi e^{i S[\psi,A]},
\end{equation}
we know that $Z_T[A]$ must be gauge invariant or equivalently have conserved $U(1)$ current $\partial_\mu \langle J^\mu\rangle_A=0$. 
But from the general anomaly inflow picture, we know that considering fermionic modes localized on the defect and the other bulk modes separately,  each would not be gauge invariant. For the localized modes current on the defect we have,
\begin{equation}
\langle J^\mu(x_1)_{\rm zm}\rangle=  -i \,e \int d^3x_2 \langle T\{J^\mu(x_1)_{\rm zm} J^\nu(x_2)_{\rm zm}\}\rangle  A_\nu (x_2),\label{eqJ}
\end{equation}
where $T\{J^\mu(x_1)_{\rm zm} J^\nu(x_2)_{\rm zm}\}$ is the time ordered correlation function of U(1) currents. Using the definition of current, $\langle J^\mu(x_1)_{\rm zm}\rangle$ will become
\begin{equation}
 i \,e \int d^3x_2 Tr(\gamma^\mu G(x_2,x_1)_{\rm zm} \gamma^\nu G(x_1,x_2)_{\rm zm}) A_\nu (x_2). \label{eqJ2}
\end{equation}
To calculate this, we need to introduce mode expansion of zero modes
\begin{align}
 \psi(x)_{\rm zm}&=\int_{0}^{\infty}\frac{dp}{2\pi} u_{\beta} \hat{a}_p \mathcal{F}(x_{\bot})  e^{-ip(t+x)}\nonumber\\
 &+\int_{-\infty}^{0}\frac{dp}{2\pi} u_{\beta} \hat{b}^\dagger_{-p}\mathcal{F}(x_{\bot})  e^{-ip(t+x)}, 
\end{align}
where $\hat{a}_p$ is the annihilation operator for the particle with momentum $p$ and $\hat{b}^\dagger_{-p}$ is the creation operator for the antiparticle with the same momentum. One can write this as
\begin{align}
 &\psi(x)_{\rm zm}=\mathcal{F}(x_{\bot}) \zeta(t,x)\label{zero mode expansion}  \\
 &=\left[\int_{0}^{\infty}\frac{dp}{2\pi}[u_{\beta} \hat{a}_p   e^{-ip(t+x)}+ u_{\beta} \hat{b}^\dagger_{p}\mathcal{F}(x_{\bot})  e^{ip(t+x)}]\right] \mathcal{F}(x_{\bot}),\nonumber    
\end{align}
where 
\begin{equation}
\zeta(t,z)=\int_{0}^{\infty}\frac{dp}{2\pi}\left[u_{\beta} \hat{a}_p   e^{-ip(t+x)}+ u_{\beta} \hat{b}^\dagger_{p} e^{ip(t+x)}\right],
\end{equation}
is a fundamentally two-dimensional chiral fermion. Now defining the Green's function as usual
\begin{align}
  &G(x_2,x_1)_{\rm zm}\\
  &= \theta(t_2-t_1) \langle \psi (x_2) \bar{\psi} (x_1)\rangle - \theta(t_1-t_2) \langle \bar{\psi} (x_1) \psi (x_2)\rangle,  \nonumber
\end{align}
We use the mode expansion~\eqref{zero mode expansion} to  obtain the Green's function of such zero modes in terms of 2D gamma matrices as subspaces of original 3D gamma matrices
\begin{align}
&G(x_2^a,x_2^\perp;x_1^a,x_1^\perp)_{\rm zm}=\nonumber\\
& \int \frac{d^2p}{(2\pi)^2} e^{i p.(x_2-x_1)}\mathcal{F}(x_2^\perp)\mathcal{F}(x_1^\perp)\frac{P_L}{\cancel{p}+i\epsilon}.    
\end{align}
where $P_L$ is 2D representation of the projection operator into left-handed chiral fermions. With this Green's function, we can calculate the one loop current of (\ref{eqJ}).  To do so, we write the spinor summations in terms of intrinsic two-dimensional gamma matrices. 
\begin{widetext}
	\begin{align} 
	\langle J^\mu(x_2)_{\rm zm}\rangle  &= \frac{i e^2}{2} \int dx_1^\perp \int\frac{d^2 p}{(2\pi)^2} \int\frac{d^2 q}{(2\pi)^2} e^{iqx_2} \mathcal{F}^2(x_2^\perp)\mathcal{F}^2(x_1^\perp)\nonumber
	 \text{Tr}\left[\gamma^\mu  \frac{P_L}{\cancel{p}+i \epsilon} \gamma^\nu  \frac{P_L}{\cancel{p}-\cancel{q}+i \epsilon}\right] A_\nu (q)
	\label{eq2J}\\
	&=\frac{i e^2}{2}  \int\frac{d^2 p}{(2\pi)^2} \int\frac{d^2 q}{(2\pi)^2} e^{iqx_2} \mathcal{F}^2(x_2^\perp)\text{Tr}\left[\gamma^\mu  \frac{P_L}{\cancel{p}+i \epsilon}\gamma^\nu  \frac{P_L}{\cancel{p}-\cancel{q}+i \epsilon}\right] A_\nu (q)\\
	&=  e^2   \int\frac{d^2 q}{(2\pi)^2} e^{iqx_2} \mathcal{F}^2(x_2^\perp) \left(\frac{ q^\mu \epsilon^{\nu \beta} q_\beta+q^\nu \epsilon^{\mu \alpha} q_\alpha}{8 \pi q^2}\right) A_\nu (q).\label{bdy current1}
	\end{align}
\end{widetext}

Using this current we can compute $\partial_\mu J^\mu$ for the pair of Dirac theories, which is proportional to the anomaly $\mathbf{a}$,
\begin{equation}
\mathbf{a}=\frac{-e^2 }{ 8\pi }   \mathcal{F}^2(x_2^\perp) \epsilon^{ab} F_{ab}.
\end{equation}
By the definition of anomaly~(\ref{eq anomaly def}), we know that the gauge variation of effective action of zero modes is proportional to the anomaly
\begin{equation}
\delta_\Lambda S_{z.m.}=\int d^3x \, \mathbf{a} \ \Lambda,
\end{equation}
which for the covariant chiral anomaly case with $\mathbf{a}=-\frac{e}{4\pi} \epsilon^{ab} F_{ab} \mathcal{F}^2(x_\perp)$, one would get
\begin{equation} \label{eq bdy1}
\delta_\Lambda S_{\rm zm}=\int d^2x dx_{\perp} \, \left(-\frac{e^2}{4\pi} \epsilon^{ab} F_{ab} \mathcal{F}^2(x_\perp)\right) \ \Lambda(x).
\end{equation}
Following~\cite{Freed1998,Harvey_2001} the gapped bulk should have effective action
\begin{equation}\label{IQH EFT}
S=-\frac{e^2}{8 \pi}\int d^3x \epsilon^{\mu\nu\lambda}(1+\rho)A_\mu F_{\nu \lambda}. 
\end{equation}
Such action fails to fulfill the Gauge invariance by the following amount:
\begin{align} \label{eq bulk1}
\delta_\Lambda S_{\rm zm}&=\left(\frac{e}{4 \pi}\right)\int d^2x dx_{\perp} \,  \epsilon^{\mu\nu\lambda}(\partial_\mu\rho) F_{\nu \lambda}\Lambda(x),\\
&=\left(\frac{e}{4 \pi}\right)\int d^2x dx_{\perp} \,  \epsilon^{\perp a b}(\partial_\perp\rho) F_{a b}\Lambda(x).
\end{align}
So by using equations (\ref{eq bdy1}) and (\ref{eq bulk1}) and considering the difference between covariant and consistent anomaly, one can choose the appropriate integration constant and see that
\begin{equation}
\rho(z)=-1+\int_{0}^{z} dz^\prime \mathcal{F}^2(z^\prime)),
\end{equation}

or explicitly
\begin{equation}
\rho(z)=-1+\mathcal{A}^2 \int_{0}^{z} dz^\prime e^{-2  \int_0^{z^\prime} g \phi(z^{\prime \prime}) dz^{\prime\prime}}.
\end{equation}
where $\mathcal{A}$ will be obtained by normalization condition (\ref{normalization1}) and depends on the specifics  of material. The function $\rho(z)$ captures the effect of zero mode profile and how effective action in the Bulk depends on it.

\section{Non-universal corrections to the Hall conductance}
\label{corrections.sec}
Our goal here is to explore the consequences of the above results. So let us consider an anomalous Hall system with a pair of Gapped Dirac particles in the background of a domain wall. Considering a pair of such Dirac particles, the current in the bulk would become
\begin{equation}
J^\mu= \frac{\delta S}{\delta A_\mu}=\frac{-e^2}{4\pi}\left(\epsilon^{\mu\nu\lambda}(1+\rho) F_{\nu \lambda}+\epsilon^{\mu\nu\lambda}(\partial_\nu\rho) A_{ \lambda}\right)
\end{equation}
which  can be explicitly  written  as

\begin{align}
J^\mu &= \frac{-e^2}{4\pi} \mathcal{A}^2  \left[ \epsilon^{\mu\nu\lambda}(\int_{0}^{z} dz^\prime e^{-2  \int_0^{z^\prime} g \phi(z^{\prime \prime}) dz^{\prime\prime}}) F_{\nu \lambda}\nonumber\right.\\
&\left.- \epsilon^{\mu\nu z}(e^{-2  \int_0^{z^\prime} g \phi(z^{\prime \prime}) dz^{\prime\prime}}) A_{ \nu} \right]
\end{align}
The second term would mainly contribute to the boundary current and is also responsible for the difference between consistent and covariant anomaly that bulk induces on the boundary whereas the first term is the Hall current in the bulk \cite{BARDEEN1984421}.

Using the normalization relation of (\ref{normalization1}), if we assume the total width of material is $L$ we  have
\begin{equation}
\mathcal{A}^{-2}=\int_{0}^{L} dz^\prime e^{-2  \int_0^{z^\prime} g \phi(z^{\prime \prime}) dz^{\prime\prime}}
\end{equation}
Then we will have the bulk currents as
\begin{equation}
J_{\rm bulk}^\mu=\frac{-e^2}{4\pi}  \left(\frac{\epsilon^{\mu\nu\lambda}F_{\nu \lambda}\int_{0}^{z} dz^\prime e^{-2  \int_0^{z^\prime} g \phi(z^{\prime \prime}) dz^{\prime\prime}}}{\int_{0}^{L} dz^\prime e^{-2  \int_0^{z^\prime} g \phi(z^{\prime \prime}) dz^{\prime\prime}}}\right). 
\end{equation}
So we can read the Hall conductance defined by 
\begin{equation}\label{hallcoefficient}
\sigma_{\rm Hall}=\lim\limits_{\omega\rightarrow 0} \frac{1}{i\omega}\langle j^x(q,\omega) j^y(-q,\omega)\rangle
\end{equation}
to be
\begin{equation}
\sigma_{\rm Hall}(z)=\frac{e^2}{2\pi} \left(\frac{ \int_{0}^{z} dz^\prime e^{-2  \int_0^{z^\prime} g \phi(z^{\prime \prime}) dz^{\prime\prime}}}{\int_{0}^{L} dz^\prime e^{-2  \int_0^{z^\prime} g \phi(z^{\prime \prime}) dz^{\prime\prime}}}\right) ,
\end{equation}
which  depends on both the value of gap controlled  $g$, and also the gap profile $\phi(z)$. 

Also, the bulk action would induce a current on the boundary as 
\begin{equation}
J_{\rm boundary}^\mu= \frac{e^2}{4\pi}\left(\frac{\epsilon^{\mu\nu z}A_{ \nu} e^{-2  \int_0^{z^\prime} g \phi(z^{\prime \prime}) dz^{\prime\prime}} }{\int_{0}^{L} dz^\prime e^{-2  \int_0^{z^\prime} g \phi(z^{\prime \prime}) dz^{\prime\prime}}}\right) ,
\end{equation}
which is not gauge invariant and it must be added to the boundary current due to boundary zero modes in equation (\ref{bdy current1}).  Note that $\epsilon^{\mu\nu z}$ essentially means a Levi-Civita symbol with two superscripts $\mu\nu$. The superscript $z$ is kept to emphasize the procedure of reduction in the dimension from the bulk to the boundary.  So for the total boundary current, we would have
\begin{eqnarray}
&&J_{\rm boundary}^\mu(x,z)=\int\frac{d^2 q}{(2\pi)^2} e^{iqx_2} \left(\frac{e^2}{2\pi}\right)\times \\
&&\left[ \frac{e^{  \int_0^{z^\prime} g \phi(z^{\prime \prime}) dz^{\prime\prime}}}{\int_{0}^{L} dz^\prime e^{-2  \int_0^{z^\prime} g \phi(z^{\prime \prime}) dz^{\prime\prime}}}  \left[ \epsilon^{\mu\nu z}  -  (\frac{\epsilon^{\mu\rho z}q^\nu q_\rho }{  q^2})  \right]A_{ \nu}(q,z)\right]\nonumber
\end{eqnarray}

Now we have the gauge invariant current on both the interface and inside the bulk. Let us consider a specific $\phi$ profile as an explicit example of this construction
\begin{equation}
\phi(z)=\tanh\left(\frac{z}{b}\right)
\end{equation}
where $b$ is a parameter that controls the width of the interface between two sections with positive and negative mass (figure \ref{wall1}) and corresponds to the localization width of the zero modes in the boundary. For such a profile we have
\begin{equation}
e^{-2  \int_0^{z^\prime} g \phi(z^{\prime \prime})}= \cosh ^{-2 g b}\left(\frac{z^\prime}{b} \right) \equiv c(z')^{-2gb}
\end{equation}
Let us compute the normalization factor in the denominator.  Note that if we consider space-filling material then we should integrate over the whole space. 

\begin{align}
&\int_{0}^{L}dz^\prime e^{-2  \int_0^{z^\prime} g \phi(z^{\prime \prime})} =	\int_{0}^{L}dz^\prime  \cosh ^{-2 g b}\left(\frac{z^\prime}{b} \right)\\
&=\frac{ \frac{\sqrt{\pi } \Gamma \left(\frac{3}{2}-b g\right)}{\Gamma (1-b g)}-c(L)^{1-2 b g} \, _2F_1\left(\frac{1}{2},\frac{1}{2}-b g;\frac{3}{2}-b g;c(L) ^2\right)}{\frac{i}{b}-2ig}\nonumber
\end{align} 
where $_2F_1$ is the Hypergeometric function. So for the bulk current, we would formally have
\begin{align}
 &J_{\rm bulk}^\mu(z)= \frac{-e^2}{4\pi}\epsilon^{\mu\nu\lambda} F_{\nu \lambda}\times\\
&\left[\frac{c(z) ^{1-2 g b} \, _2F_1\left(\frac{1}{2},\frac{1}{2}-g b;\frac{3}{2}-g b;c(z)^2\right)-\frac{\sqrt{\pi } \Gamma \left(\frac{3}{2}-g b\right)}{\Gamma \left(1-g b\right)}}{c(L) ^{1-2 g b} \, _2F_1\left(\frac{1}{2},\frac{1}{2}-g b;\frac{3}{2}-g b;c(L)^2\right)-\frac{\sqrt{\pi } \Gamma \left(\frac{3}{2}-g b\right)}{\Gamma \left(1-g b\right)}}\right]\nonumber .
\end{align}
Therefore the Hall conductance (\ref{hallcoefficient})  will become 

\begin{align}
&\sigma_{\rm H}(z)= \frac{e^2}{2\pi}\times\\
&\left[\frac{c(z)^{1-2 g b} \, _2F_1\left(\frac{1}{2},\frac{1}{2}-g b;\frac{3}{2}-g b;c(z)^2\right)-\frac{\sqrt{\pi } \Gamma \left(\frac{3}{2}-g b\right)}{\Gamma \left(1-g b\right)}}{c(L)^{1-2 g b} \, _2F_1\left(\frac{1}{2},\frac{1}{2}-g b;\frac{3}{2}-g b;c(L)^2\right)-\frac{\sqrt{\pi } \Gamma \left(\frac{3}{2}-g b\right)}{\Gamma \left(1-g b\right)}} \right].
\end{align}
So instead of a constant Hall conductance across the material, we have $z$ dependent conductance, depending on both the width of boundary and bulk and also the value of gap via $g$.  The entire measured current in the boundary is obtained by integrating over the whole geometry. Assuming Hall bar geometry we have
\begin{equation}
\sigma_{\rm H}=\frac{1}{L}\int_0^L \sigma(z) dz.
\end{equation}
In order to write the above formula in the standard units, one notes that $g$ has the dimension of energy. Therefore when combined with $b$ of dimension length, will have the dimension of $\hbar v_F$ where $v_F$ is the (Fermi) velocity scale of the material and $\sigma_{\rm H}$ will be expressed in the natural unit of $e^2/h$. Doing so to restore the constants we obtain
\begin{widetext}
\begin{equation}
\sigma_{\rm H}= \frac{e^2}{h }\int_0^L \frac{dz}{ L}   \left[\frac{c(z)^{1-2 \frac{g b}{\hbar v_F}} \, _2F_1\left(\frac{1}{2},\frac{1}{2}-\frac{g b}{\hbar v_F};\frac{3}{2}-\frac{g b}{\hbar v_F};c(z)^2\right)-\frac{\sqrt{\pi } \Gamma \left(\frac{3}{2}-\frac{g b}{\hbar v_F}\right)}{\Gamma \left(1-\frac{g b}{\hbar v_F}\right)}}{c(L)^{1-2 \frac{g b}{\hbar v_F}} \, _2F_1\left(\frac{1}{2},\frac{1}{2}-\frac{g b}{\hbar v_F};\frac{3}{2}-g b;c(L)^2\right)-\frac{\sqrt{\pi } \Gamma \left(\frac{3}{2}-\frac{g b}{\hbar v_F}\right)}{\Gamma \left(1-\frac{g b}{\hbar v_F}\right)}} \right]
\label{finalSigmaH.eqn}
\end{equation}
\end{widetext}
where $\hbar$ is reduced Planck constant and $v_F$ is Fermi velocity. Using the above formulas one can find the Hall conductance for different setups. We see that there are two dimensionless parameters controlling the quantization of Hall coefficient: One parameter is $\frac{g b}{\hbar v_F}$ which depends on the value of gap $(g)$, localization length of edge modes $(b)$, Fermi velocity of material $(v_F)$ and the other parameter is $\frac{b}{L}$ or the ratio of edge localization and the sample length.

\section{Sample dependent deviations from $e^2/h$}
\label{results.sec}

\begin{figure}[t] 
	\centering
	\includegraphics[width=0.5\textwidth]{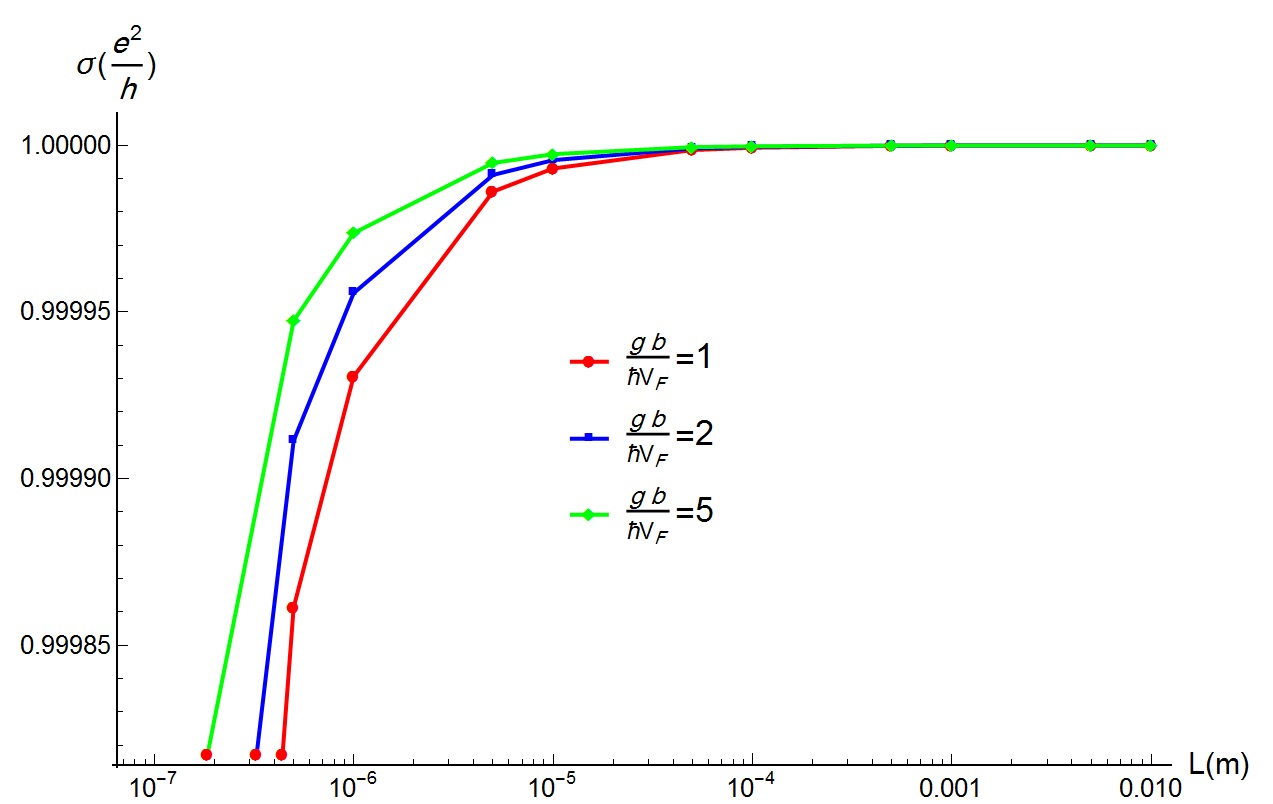}
	\caption{Dependence of Hall coefficient on sample length $L$ for a atomically localized edge modes with $b=10^{-10}m$}
	\label{hallL}
\end{figure}
Let's observe how by fixing two parameters and changing the other one, the Hall coefficient would change. we probe the values that are relevant to the observed experimental realizations of quantum anomalous hall experiments \cite{Chang2013,PhysRevLett.114.187201,PhysRevLett.113.137201,Kou_Pan_Wang_Fan_Choi_Lee_Nie_Murata_Shao_Zhang_etal._2015}. 

\begin{figure}[b] 
	\centering
	\includegraphics[width=0.5\textwidth]{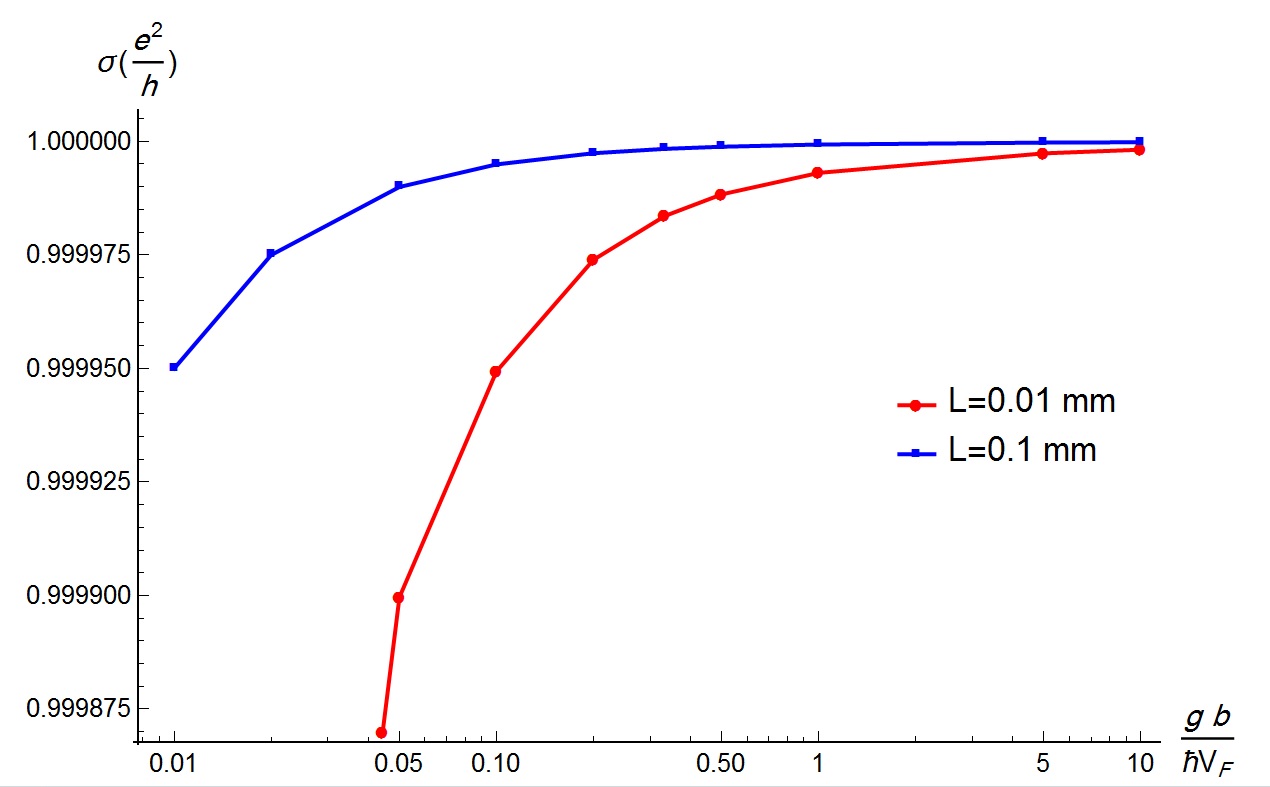}
	\caption{Hall coefficients as a function of gap value $g$ for a material with atomically localized edge modes with $b=10^{-10}m$}
	\label{hallg}
\end{figure}
As can be seen in Fig.~\ref{hallL} fixing both the edge mode localization length $b$ to a typical atomic size of $1\AA$, and the ratio of $\frac{g}{v_F}$ when the length of the material $L$ drops below $mm$  length scale, deviations from the universal value $e^2/h$ become manifest at the fourth digit. As can be seen for typical values of the gap $gb/(\hbar v_F)$ shown in the figure, the deviations are stronger for smaller $gb/(\hbar v_F)$ values. For larger values of the gap, the deviations from the universal value sets in at lower sample length scales. This indicates that the larger bulk gaps are closer to the universal limit. In fact, in the limit of infinitely large gap the conductance will have its universal value, irrespective of the sample size $L$. 

The dependence on the gap can be more manifestly seen in Fig.~\ref{hallg} where for two samples of length $0.1$mm and $0.01$ mm the dependence of the Hall conductance on gap has been depicted. As can be seen for larger gap values the Hall conductance tends to the inverse von Klitzing constant. 

\begin{figure}[t] 
	\centering
	\includegraphics[width=0.5\textwidth]{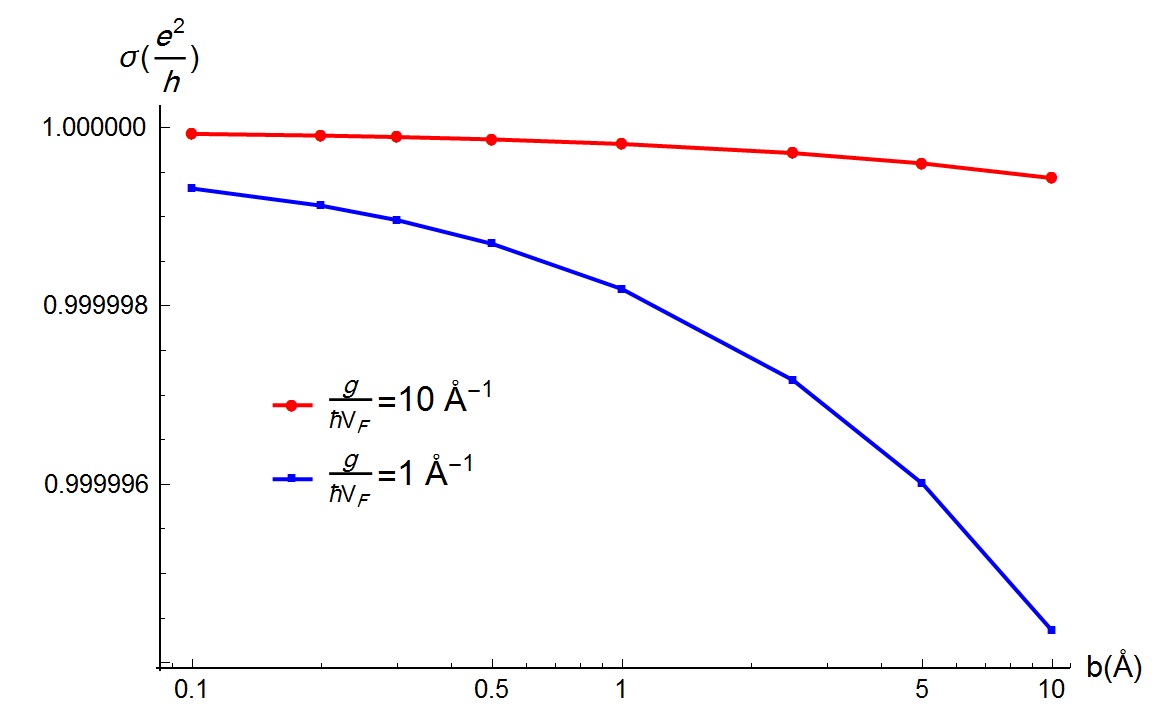}
	\caption{Hall coefficients with respect to changing localization value b for material with total length of 0.1 mm}
	\label{hallb}
\end{figure}

Finally, in Fig.~\ref{hallb} we have depicted the variations of the Hall conductance as a function of the localization length $b$ for a fixed value of $L=0.1$ mm and fixed values of $\frac{g}{v_F}=1, 10\AA^{-1}$ with blue and red curves, respectively. The blue curve corresponds to gap values on the scale of a fraction of eV if $b$ is assumed to be a fraction of $\AA$ which can be relevant to Cr doped Bi based topological insulators~\cite{Chang_Zhang_Feng_Shen_Zhang_Guo_Li_Ou_Wei_Wang_etal}.

It is important to note that localization length $b$, affects the Hall coefficient through two competing tendencies, one as the exponent of $\cosh$ and the other one as argument of the hypergeometric function. Therefore the behavior under variation of $b$ is more complicated. Although in general, decreasing the value of edge mode localization length would force the Hall coefficient to get closer to its universal value.

\section{Summary and discussions}
\label{discussions.sec}
In this work, we have used the anomaly inflow framework to compute the corrections to Hall conductance in QAHE that arise from a finite localization length $b$ for the boundary zero modes. In the limit where $b$ is zero, i.e. the edge modes are sharply (Dirac delta) localized in the boundary, the Hall conductance as expected is given by its universal value of $e^2/h$ -- the inverse von Klitzing constant. As soon as $b$ becomes non-zero, quite unexpectedly the Hall conductivity's dependence not only on the dimensionless ratio $b/L$ where $L$ is the Hall bar length sets in, but also a dependence on another dimensionless variable $\frac{gb}{\hbar v_F}$ is triggered where $g$ is the bulk gap and $v_F$ is the Fermi velocity. Therefore not only the sample size start to non-universally contribute to the Hall conductance, but also the composition of the material affects the Hall conductance via the ratio $g b/\hbar v_F$ for any given localization length $b$ of the edge modes. 

 In the $1\%$ accuracy of the Ref.~\cite{Bestwick2015} dominant thermal activation across the bulk gap can mask smaller boundary localization effects discussed here. The observed high precision Hall conductance of $(0.9998\pm0.0006)e^2/h$  in hard ferromagnetic topological insulators~\cite{Chang2015} might be a hint for significant boundary localization effects. 

Our results show how abstract theoretical notions of anomaly inflow can find concrete condensed matter realization in the context of QAHE. Our theory further offers a framework to "measure" the localization length $b$ by fitting the accurately measured Hall conductance for a handful of sample lengths of a given material where the ratio $g/v_F$ is fixed. The cross over from edge-dominated transport in experiments of Ref.~\cite{Kandala2015} to diffusive transport in the bulk as a function of tilting the field angle can be thought of as a means for tuning the gap $g$ where upon tilting the field away from perpendicular direction, the gap becomes smaller, allowing for deviations from the universal quantization. Combining high precision measurements of the type~\cite{Chang2015} at low-temperatures with tilted field studies of the type~\cite{Kandala2015} where the gap $g$ can be tuned to very small values when the thermal effects~\cite{Bestwick2015} are carefully separated can be a route to reveal the anomaly inflow physics discussed here that manifests itself as non-universal corrections to the Hall conductance.

\section{Acknowledgements}
This work is part of the requirements for the degree PhD for A.G. and was conducted under the supervision of S.A.J while he was a professor at Sharif University of Technology. 
\newpage

\bibliography{bibtex}



\end{document}